\documentclass{article}

\usepackage{styles}
\usepackage{babel}

\usepackage[utf8]{inputenc} 
\usepackage[T1]{fontenc}    
\usepackage{hyperref}       
\usepackage{url}            
\usepackage{booktabs}       
\usepackage{amsfonts}       
\usepackage{nicefrac}       
\usepackage{microtype}      
\usepackage{changepage}     
\usepackage{float}          
\usepackage{lipsum}
\usepackage{graphicx}
\usepackage{amsmath,amssymb,amsfonts}
\usepackage{booktabs}
\graphicspath{ {./images/} }
\usepackage{natbib}
\usepackage{tikz}
\usepackage{tabularx}
\usepackage{array}
\newcolumntype{C}{>{\centering\arraybackslash}X}
\usetikzlibrary{positioning}
\title{Time Entangled Quantum Blockchain with Phase Encoding for Classical Data}

\author{
 Ruwanga Konara \\
  School of Computing \\
  University of Colombo\\
  Colombo, Sri Lanka \\
  \texttt{ruwangathandulakonara@gmail.com} \\
   \And
 Kasun De Zoysa \\
School of Computing \\
  University of Colombo\\
  Colombo, Sri Lanka \\
  \texttt{kasun@ucsc.cmb.ac.lk} \\
  \And
 Anuradha Mahasinghe \\
  Department of Mathematics\\
  University of Colombo\\
  Colombo, Sri Lanka \\
  \texttt{anuradhamahasinghe@maths.cmb.ac.lk} \\
  \And
   Asanka Sayakkara \\
School of Computing \\
  University of Colombo\\
  Colombo, Sri Lanka \\
  \texttt{asa@ucsc.cmb.ac.lk} \\
    \And
   Nalin Ranasinghe \\
School of Computing \\
  University of Colombo\\
  Colombo, Sri Lanka \\
  \texttt{dnr@ucsc.cmb.ac.lk} \\
}

\newlength{\extralength}
\begin{document}
\maketitle
\begin{abstract}
Rapid progress in quantum computing threatens the long‑term security of classical cryptographic primitives, and with them the integrity of contemporary blockchain systems that rely fundamentally on computational hardness assumptions. Hence, quantum‑native blockchain architectures have emerged as a conceptual pathway toward information‑theoretic disturbance detectability. Two influential approaches have emerged in the literature. The temporal GHZ-state blockchain provides disturbance-detectable tamper sensitivity through entanglement in time, whereas the weighted quantum-hypergraph blockchain achieves high encoding efficiency through phase-based quantum representations of classical information. However, each addresses only part of the problem. In this work, we introduce a hybrid quantum blockchain framework whose primary novelty is the integration of phase-encoded classical data representation with recursively generated temporal GHZ entanglement within a single blockchain architecture. Rather than proposing a new encoding scheme or a new temporal-entanglement construction, the framework combines both mechanisms found in the literature and introduces a corresponding verification procedure for validating phase-encoded temporally entangled blocks. This architecture preserves the physics‑based measurement‑disturbance detectability of temporal entanglement while enabling more efficient classical‑to‑quantum data encoding inspired by hypergraph‑based phase weighting. The result is a conceptual blockchain model that simultaneously enhances tamper sensitivity and encoding efficiency, providing a coherent foundation for future research on secure and practical quantum‑era ledger systems.
\end{abstract}

\keywords{Quantum blockchain \and Physical tamper sensitivity \and Disturbance detectability \and Time Entanglement \and Phase \and Quantum Hypergraph \and GHZ State \and Consensus}

\section{Introduction}\label{intr}

subsection{Classical Blockchain}

A blockchain is an immutable ledger where data is stored on data structures known as blocks connected by hash pointers generated in chronological order. It is essentially a trustless peer-to-peer network \citep{Nakamoto2008Bitcoin}, where time-stamped information (transactions in a cryptocurrency) is drawn from a pool of transactions and stored as a merkle tree in a block by a block creator: one of the nodes of the network chosen via a consensus protocol. These data are time-stamped (and encrypted if necessary) records: data about the past. It is a decentralized network that facilitates transactions/data without the necessity of a central party to authorize/authenticate its data and processes. Validated timestamped records that were put into the network in a given period by network nodes are packaged into a block and appended to the previous period's block via a hash function \citep{Fartitchou2022}. The new block stores the hash of the previous block. The most famous use case of blockchain is Bitcoin \citep{Nakamoto2008Bitcoin}. A blockchain stores data on which all nodes have reached a consensus (agreement). There are many such consensus protocols, among which the most popular is PoW (Proof of Work). A block creator chosen by consensus validates the transactions he has put into the block and sends the block to the network where nodes validate the block and its transactions and append it to their local copy of the blockchain. This technology has found usage in various fields such as medicine \citep{lin2023} and social media \citep{esmeralda2022}. The decentralized nature and the links formed by hash functions between blocks make it computationally infeasible for an adversary to mutate a past record. That is, if a block is mutated, its hash has to be recalculated and changed in the following block. And the adversary would recursively have to recalculate all the hashes for blocks following the mutated block. Also, since all nodes have a copy of the chain, the adoption of the mutated chain must be ensured as well. This requires the attacker to control more than half of the hash rate of the network, thus creating the longest local chain. The longest chain is eventually adopted by the network. Therefore, tampering with a block, due to the high degree of confidence of hash functions, effectively invalidates the blocks after the block in question. Therefore, the older the timestamp on a record in a blockchain, the safer it is: \citep{Jain2024}

\subsection{Quantum Blockchain}

With the advent of quantum computing, algorithms such as Shor \citep{shor1994} and Grover \citep{Preston2022} have shown themselves to be capable of compromising classical cryptography and hence, classical blockchain. There are applications and extensive research on quantum-immune classical blockchains based on quantum-immune classical cryptography. \citep{FernandezCarames2020}. However, some of these schemes are broken by classical approaches and continuous research happens on new quantum algorithms to break these schemes as well. Therefore a sensible approach is to use quantum technologies to implement technical aspects of blockchain, the pinnacle of which is to define and store blockchains on a quantum register. These are known as quantum blockchains. An extensive analysis of quantum blockchains is in \citep{Jain2024}. There are two main conceptualizations of quantum blockchains. The first is Del Rajan and Matt Visser's \citep{rajan2019} quantum blockchain on the temporal GHZ\footnote{A GHZ state where particles are entangled not only in space but also in time: \cite{rajan2020quantum}.} (Greenberger-Horne-Zeilinger) state. The other is Shreya Banerjee's \citep{shreya2020} blockchain on a quantum hypergraph. The GHZ blockchain has physics-based disturbance detectability 
in its temporal entanglement, while the hypergraph blockchain shows encoding efficience\footnote{the time-entangled GHZ blockchain encodes a classical block of 2 bits to a quantum block of 2 qubits. Our framework is more efficient in the sense that a classical block of arbitrary size is mapped to a phase parameter and auxiliary classical metadata (a two-bit classical string), which are then represented using a two-qubit Bell-state block.

However, not as encoding efficient as the quantum hypergraph blockchain \citep{shreya2020}, which encodes a classical block of arbitrary size by mapping it to a phase parameter, which is represented by a single qubit} since, in their scheme, a quantum block is equivalent to the data stored in an entire classical block. We conceptualize a new blockchain architecture with both these properties combining features of both these conceptualizations. The structure of this work is as follows. The GHZ blockchain and the quantum hypergraph blockchain are explained. Then the tamper evidence (disturbance detectability) details of the GHZ blockchain and the efficiency of the hypergraph blockchain are outlined. The proposed quantum blockchain framework is described: the ledger data structure, quantum network, security definitions, consensus, and authentication. The section beyond that discusses future work to be done and comes before the concluding chapter.

The core contribution of this work is the integration of temporally entangled GHZ blockchain's tamper sensitivity with quantum hypergraph blockchain-inspired phase encoding\footnote{Mapping the classical block into a phase angle and adding that phase to Bell and GHZ states.}, forming a unified quantum blockchain model that simultaneously addresses tamper sensitivity and encoding efficiency. Security claims in this work refer specifically to measurement-disturbance detectability and state-integrity disruption under direct quantum access attacks, not to a complete cryptographic security model. Beyond the two blockchain frameworks in the literature that inspired the current work, we discuss other recent research that is instrumental in positioning our work.

\subsection{Other Related Work}

Recent work has expanded quantum blockchain research beyond the two early reference models used in this manuscript. Surveys have classified the field into post-quantum blockchains and fully quantum blockchains \citep{Yang2024Survey,Khodaiemehr2023Navigating}. These surveys outline quantum threats to classical blockchain mechanisms, synthesize quantum blockchains \& post-quantum blockchains in the literature, and perform comparative analysis. There are application specific quantum blockchains as well: for instance, identity management \citep{Yang2022Decentralization}. Other quantum blockchain/communication adjacent work related to our work from an engineering perspective include Quantum native consensus mechanisms \citep{Li2022Efficient}, quantum blockchain frameworks for 6G and post-quantum data security \citep{Farouk2025Design,Reddy2025Quantum}, high-dimensional time-entangled blockchain proposals \citep{Aktas2025High}, and implementation improvements to hypergraph-based quantum blockchains \citep{Orts2023Improving}. Zawadzki \citep{Zawadzki2023Insecurity} is especially relevant because it questions the security interpretation of blockchains based on entanglement in time; this motivates the cautious wording we have utilized in the present work, where claims are limited to disturbance detectability under explicit assumptions.

This work is also related to broader quantum-information experiments and protocols involving tomography, Bell tests, phase encoding, and tamper detection in the sense that those advances are instrumental in the practical realization of our framework. Recent optical demonstrations of Bell tests and quantum-state tomography show the wider experimental community concerned with advancement in verifying quantum correlations \citep{Arbel2025Optical,Sanz2024Undergraduate} within the established literature on quantum state verification, property (entenglement, etc.) verification, and special state classes verification. Recent Clauser–Horne–Shimony–Holt (CHSH) tests for optical hybrid entanglement further illustrate active work on experimentally characterizing nonclassical correlations \citep{Moradi2024CHSH}. Phase encoding also appears in quantum private query and quantum secret-sharing protocols \citep{Liu2025QKD,Shen2023Experimental}, and probabilistic quantum-state models have been explored for classical data security \citep{Hafiz2023A}. Finally, recent work on quantum tamper detection \citep{Broadbent2025Towards} provides a useful theoretical context for the disturbance-detection perspective adopted here, although the present framework does not claim to realize universal quantum tamper-detection codes.

\paragraph{Scope and Assumptions}

The proposed framework should be interpreted as a conceptual blockchain architecture rather than a near-term implementation proposal. Its operation relies on several non-trivial assumptions, including stable phase control, temporal coherence, high-fidelity recursive fusion operations, synchronization of temporal modes, low decoherence, and reliable quantum-network infrastructure. These assumptions are discussed in detail in \ref{assume} and should be regarded as engineering challenges rather than experimentally fully realized capabilities.

\subsection{Contributions and Novelty}

In light of the related work, the novelty of this work is not the independent introduction of temporal entanglement or phase encoding, since these ideas have previously appeared separately in the literature. Instead, the contribution of this work is their integration into a single blockchain architecture. Also, the novelty is not in quantum consensus or quantum-network security.

Specifically, this work contributes:

\begin{itemize}

\item a hybrid blockchain architecture combining temporal GHZ entanglement with phase-based classical data encoding inspired by weighted hypergraph blockchain constructions;

\item a new variant of a method from the literature for deriving phase-based quantum representations (quantum blocks) of arbitrary-sized classical blocks before recursive temporal fusion; this original method was from \cite{shreya2020}. Our variant maps a classical block to a phase parameter and verification metadata, i.e., the two-bit string, which are subsequently represented by a quantum state.

\item a verification procedure for validating phase-encoded temporally entangled quantum blocks;

\item conceptual analysis of disturbance detectability, scalability, feasibility constraints, and consensus requirements for such a hybrid architecture.

\end{itemize}

Therefore, the primary contribution of this work is the combination of phase encoding \& temporal entanglement and verification definition within a unified blockchain architecture rather than the introduction of any individual component in isolation.

\section{Time Entangled GHZ State Blockchain } \label{rajan}

This blockchain framework was introduced in \citep{rajan2019} by Del Rajan and Matt Visser. Entanglement is a phenomenon in which a compound state cannot be presented as a tensor of the states of the particles involved. Entanglement is the non-classical correlation between spatially distant particles that was referred to as "spooky action at a distance" \citep{einstein1935can} by Einstein. It is the foundation of quantum communication such as teleportation \citep{bennet1993} and superdense coding. Progress toward a quantum internet \citep{Luo2024} is being made on these technologies and quantum phenomena. An entangled quantum state is a compound state that cannot be given as a tensor of the states of quantum particles involved. A bipartite entangled state \(|\psi_{ab}\rangle\) of particles \(a\) and \(b\) conforms to the following

\begin{equation}
|\psi_{ab}\rangle \neq |a\rangle \otimes|b\rangle
\end{equation}

where \(|a\rangle\) and \(|b\rangle\) are individual qubit states. 

 multipartite GHZ (Greenberger–Horne–Zeilinger)states\citep{Carvacho2017, greenberger2007goingbellstheorem} are states in which all subsystems (particles) contribute to the shared entangled property. This is used to conceptualize a chain. A concept from superdense coding where a classical 2-bit string \(xy\) is encoded into a bell state by the following.

\begin{equation}
|B_{xy}\rangle = \frac{1}{\sqrt{2}}(|0\rangle|y\rangle + (-1)^x|1\rangle|\tilde{y}\rangle)  
\end{equation}

In this scheme, a classical block is two bits. The encoding procedure converts each classical block \(r_1r_2\), into a temporal Bell state \citep{Megidish_2013}, generated at a particular time. At \(t = 0\), the temporal Bell state (quantum block) generated is as follows:

\begin{equation}
|B_{r_1r_2}\rangle^{0,\tau} = \frac{1}{\sqrt{2}}(|0^0\rangle|r_2^\tau\rangle + (-1)^{r_1}|1^0\rangle|\tilde{r}_2^\tau\rangle)
\end{equation}

The superscript in the kets is the time at which the photon is absorbed; this provides a timestamp for the block. The authors of \citep{Megidish_2013} experimentally generated such temporal Bell states.  They presented spatial bell states with polarized photons where \(v_a (h_a)\) represent the vertical (horizontal) polarization in spatial mode \(a (b)\). To create the temporally entangled states, consecutive pairs of spatially entangled pairs are generated at defined intervals separated by interval \(\tau\).

\begin{equation}
|\psi^-\rangle_{a,b}^{0,0} \otimes |\psi^-\rangle_{a,b}^{\tau,\tau} = (|h^0_a \,v^0_b\rangle - |v^0_a\, h^0_b\rangle) \otimes (|h^\tau_a\, v^\tau_b\rangle - |v^\tau_a\, h^\tau_b\rangle)
\end{equation}

A delay line was added to one photon of each pair in \citep{Megidish_2013}.

\begin{align}
|\psi^-\rangle_{a,b}^{0,\tau}\, |\psi^-\rangle_{a,b}^{\tau,2\tau}
&=
\frac{1}{2}\Big(
|\psi^+\rangle_{a,b}^{0,2\tau}\,|\psi^+\rangle_{a,b}^{\tau,\tau}
-
|\psi^-\rangle_{a,b}^{0,2\tau}\,|\psi^-\rangle_{a,b}^{\tau,\tau}
\nonumber\\
&\quad
-
|\phi^+\rangle_{a,b}^{0,2\tau}\,|\phi^+\rangle_{a,b}^{\tau,\tau}
+
|\phi^-\rangle_{a,b}^{0,2\tau}\,|\phi^-\rangle_{a,b}^{\tau,\tau}
\Big).
\end{align}

 The photons absorbed at \(t=0\) and \(t = 2\tau\) are entangled when Bell projection is carried out on two photons at \(t = \tau\). The former pair of photons never even coexists. In the design of the ledger data structure in \citep{rajan2019}, 2-bit blocks are encoded into temporal Bell states, where photons are absorbed at respective times of encoding. The first three blocks,

 \[
|B_{00}\rangle^{0,\tau}, \quad|B_{10}\rangle^{\tau,2\tau}, \quad|B_{11}\rangle^{2\tau,3\tau}
\]

These states (blocks) are chained together through entanglement in time. This is done by recursively projecting these Bell states into a growing GHZ state an entanglement source, a delay line, and a polarizing beam splitter (PBS). The following is an example of two Bell pairs forming a four qubit GHZ state

\begin{align}
|\psi^{+}\rangle_{a,b}^{0,0} \otimes |\psi^{+}\rangle_{a,b}^{\tau,\tau}
&\overset{delay}{\longrightarrow}
|\psi^{+}\rangle_{a,b}^{0,\tau} \otimes |\psi^{+}\rangle_{a,b}^{\tau,2\tau}
\nonumber\\
&=
\frac{1}{2}
\left(
|h_{a}^{0}\,v_{b}^{\tau}\rangle
+
|v_{a}^{0}\,h_{b}^{\tau}\rangle
\right)
\otimes
\left(
|h_{a}^{\tau}\,v_{b}^{2\tau}\rangle
+
|v_{a}^{\tau}\,h_{b}^{2\tau}\rangle
\right)
\nonumber\\
&\overset{PBS}{\longrightarrow}
\frac{1}{2}
\left(
|h_{a}^{0}\,v_{b}^{\tau}\,h_{a}^{\tau}\,h_{b}^{2\tau}\rangle
+
|v_{a}^{0}\,h_{b}^{\tau}\,h_{a}^{\tau}\,v_{b}^{2\tau}\rangle
\right)
\nonumber\\
&=
|GHZ\rangle_{0,\tau,2\tau}.
\end{align}

According to \citep{rajan2019}, "Entanglement exists between the four photons that propagate in different spatial modes and exist at different times." From \(t = 0\), the state of the blockchain at \(t = n\tau\):

\begin{equation}
    |GHZ_{r_1 r_2 \dots r_{2n}}^{0,\tau,\tau,2\tau,2\tau \dots,(n-1)\tau,(n-1)\tau,n\tau}\rangle = \frac{1}{\sqrt{2}}(|0^0\, r_2^\tau\, r_3^\tau\, \dots\, r_{2n}^{n\tau}\rangle + (-1)^{r_1}|1^0\, \bar{r}_2^\tau\, \bar{r}_3^\tau\, \dots\, \bar{r}_{2n}^{n\tau}\rangle)
\end{equation}

Subscripts on the left side denote the concatenated string of all the blocks, i.e all the data stored on the chain. At \(t = n\tau\), only one photon remains. The following is to quote \citep{rajan2019}.

\begin{quote}
    It is important to note that at this stage of development, we are advocating a conceptual mathematical design for a new quantum information technology. It should be viewed as analogous to early quantum algorithms (Deutsch’s algorithm \citep{Deutsch1985Quantum}, Deutsch-Jozsa algorithms \citep{deutsch1992}). In the 1980s, the engineering considerations for quantum computers were not taken into account. In the 1990s, when Shor’s algorithm and Grover’s algorithm were developed, the experimental realization of quantum computers was almost seen as an impossible project. Nonetheless, their work was certainly of interest to the wider community \citep{needle1997}.
\end{quote}

\section{Quantum Blockchain on a Weighted Hypergraph }\label{shreya}

This scheme was presented in \citep{shreya2020} by Shreya Banerjee, A. Mukherjee, and P. K. Panigrahi. Quantum hypergraphs \citep{Rossi_2013} are a group of highly entangled multiparty quantum states. Vertices are quantum particles where edges represent entanglement. Edges encompass more than two vertices and are known as "hypergraphs." With a quantum hypergraph with \(k\)-hyperedge (a hyperedge connecting \(k\) qubits) and \(n -1\) vertices, a corresponding quantum state can be prepared \citep{Rossi_2013}. To add the \(n\)th qubit to the state, initialize the qubit to the Hadamard state: \((|0\rangle + |1\rangle)/\sqrt{2}\). A controlled \(Z\) operation is performed with the existing \(n - 1\) qubits as control and \(n\) as the target. For a mathematical hypergraph with ﬁve vertices, A, B, C, D, E, a 3-hyperedge over vertices A, B, and C; and a 5-hyperedge over all ﬁve vertices, the state is expressed as following.

\begin{equation}
|\psi\rangle = C^2_{(A, B, C, D, E)}Z C^2_{(A, B, C)}Z |+\rangle^{\otimes 5}    
\end{equation}

The entanglement of a hypergraph state and its properties have been discussed in \citep{Gühne_2014, Rossi_2013, Qu2013}. 

\citep{Gühne_2014} describes how local unitary operations carried through classical communication (LOCC) does not alter the entanglement of the state under consideration. It is shown in \citep{Gühne_2014} that local applications of unitary Pauli operations on the \(n\)th qubit one can remove all the \((N - 1)\) edges for the special case where an \(n\)-hypergraph (\(n\)-qubit hypergraph) contains only an \(n\) hyperedge. As presented in \citep{Qu2013, Tsimakuridze_2017}, a weighted hypergraph is when hyperedge carries weights. \citep{Qu2013, kruszy2009} introduced weighted quantum hypergraphs as locally maximally entangleable (LME) states. In the notation below, \(|x\rangle\) is the computational basis, and \(f(x)\) is a real number.

\begin{equation}
|\psi\rangle = \frac{1}{\sqrt{2^N}} \sum_{x \in (0,1)^N} e^{i\pi f(x)} |x\rangle.    
\end{equation}

In the proposed blockchain, the classical ledger and cryptographic functions have been replaced by the weighted quantum hypergraph and entanglement. They have used the phases (weights) on the hyperedges to encode classical information. Each classical block is a string \(P_i\) (\(i^{th}\) classical block) of bits. The block creator initializes the qubit (one qubit represents one quantum block) to \((|0\rangle + |1\rangle)/\sqrt{2}\). Then via a secret bijective function chosen known only by him, \(P_i\) is converted into the phase \(\theta_{P_i}\), and the following rotation is applied to the qubit to obtain the \(i^{th}\) quantum block. 

\begin{equation}
    |\psi_i\rangle = S(p) |\psi\rangle = \frac{1}{\sqrt{2}}
\begin{bmatrix}
1 & 0 \\
0 & e^{i\theta_p}
\end{bmatrix}
|\psi\rangle = \frac{|0\rangle + e^{i\theta_p} |1\rangle}{\sqrt{2}}
\end{equation}

\(\theta_{P_i} \in (0, \pi/2)\) is \(f_i(P)\) the output of the aforementioned bijective function \(f_i\). \(\theta\) should satisfy \(\sum_i^\infty p_i < \pi/2\). The qubit \(|\psi_i\rangle\) (quantum block \(i\)) now carries the information in classical n block \(P_i\). Conditions on the phase are of critical importance as they ensure the entanglement of the hypergraph. These conditions are part of the consensus. Constraints include \(\theta_{P_i} = \theta_{P_1}/2^{(i-1)} \). where \(\theta_{P_1}\) is the phase of the first block. The infinite sum of all the phases 

\begin{equation}
\sum_{i=1}^{\infty} \theta_{p_i} = \sum_{i=1}^{\infty} \frac{1}{2^{i-1}} \theta_{p_1}.    
\end{equation}

converges to \(2\theta_{P_1}\). Therefore, to ensure \(\sum_i^\infty p_i <  \frac{\pi}{2}\), \(\theta_{P_1} \) needs to be less than \(\frac{\pi}{4}\).

There can be many such series.

\begin{equation}
\sum_{i=1}^{\infty} \theta_{p_i} = \sum_{i=1}^{\infty} \frac{1}{n^{i-1}} \theta_{p_1}.    
\end{equation}

where \( n \in \mathbb{N} \setminus \{1\} \). The series converges to \(\frac{n}{n-1}\). Consensus can therefore be defined with any such series with an appropriate \(\theta_{P_1} \). After the consensus execution, peers add the \(n\)th block to their local chain of \(n-1\) blocks. A \(C^{(n-1)}Z\) is applied as before with the new block as the target to create the new \(n\)-hyperedge and a local Pauli-\(X\) on the new block to remove the previous \(n-1\)-hyperedge.

\section{Hybrid Quantum Blockchain Framework}\label{new chain}

\subsection{Security and Scalability}

The GHZ blockchain \citep{rajan2019} exhibits disturbance detectability under the assumed measurement model due to the sensitivity of multipartite GHZ (maximally entangled) states to unauthorized measurements. In the spatial GHZ case, measurements performed on constituent photons modify correlation structures and may invalidate verification outcomes. The temporal GHZ blockchain further restricts access to earlier entangled components because previous temporal states are no longer simultaneously accessible. An adversary may still attempt measurements on remaining states; however, such operations modify the entangled system and may be detectable during verification. Therefore, temporal entanglement provides disturbance detectability and tamper evidence under the assumed measurement model rather than information-theoretic security guarantees. See \ref{threat} for the current threat model. The scalability of these temporal GHZ states was considered in \citep{Megidish_2013}. They have stated 

\begin{quote}
	any number of photons are generated with the same setup, solving the scalability problem caused by the previous need for extra resources. Consequently, entangled photon states of larger numbers than before are practically realizable.
\end{quote}

The scalability analysis of our protocol is in \ref{scaling}.

\paragraph{Encoding Interpretation}

The proposed framework does not claim that an arbitrary classical block can be losslessly compressed into two qubits in the Shannon-information sense. Instead, inspired by the phase-encoding philosophy of the weighted hypergraph blockchain \citep{shreya2020}, a classical block is mapped to a phase parameter together with auxiliary classical two-bit string used during verification. The two-qubit Bell-state block therefore represents the encoded block identifier rather than a complete quantum storage of all original classical information. Consequently, the scalability discussion in \ref{scaling} concerns the size of the quantum blockchain representation rather than lossless quantum compression of arbitrary classical data.

Accordingly, the present framework should be interpreted as representing a cryptographic identifier derived from the classical block, rather than storing the entire block contents within a two-qubit quantum state.

\subsubsection{Scalability Analysis} \label{scaling}

Scalability in this work refers primarily to the amount of quantum information required to represent a given amount of classical ledger data, i.e., encoding efficiency when encoding a classical block to a quantum block.

A classical block contains \(m\)bits. In the temporal GHZ blockchain \citep{rajan2019}, quantum block size scales with the amount of classical information represented. Therefore \(Q_{GHZ}(m)=O(m)\), where \( Q_{GHZ}(m)\) denotes the number of qubits required to represent a classical block containing \(m\) bits.

In the weighted quantum hypergraph blockchain \citep{shreya2020}, an arbitrary-sized classical block is mapped to a phase parameter and represented on a single qubit. Therefore, \(Q_{HG}(m)=1\), which implies \(Q_{HG}(m)=O(1)\), independent of classical block size. In the proposed framework, an arbitrary-sized classical block is mapped into a phase value \& a 2-bit string as metadata. These two items are then represented by a Bell state consisting of two qubits.

Therefore, \(Q_{Hybrid}(m)=2\), which similarly gives \(Q_{Hybrid}(m)=O(1)\) with respect to classical block size. Consequently:

\begin{equation}
   Q_{HG}(m) < Q_{Hybrid}(m) \ll Q_{GHZ}(m)
\end{equation}

for sufficiently large \(m\). This implies that the proposed framework sacrifices some encoding efficiency relative to the hypergraph blockchain, that is \( 1 \text{ qubit } \rightarrow 2 \text{ qubits}\), in exchange for stronger multipartite entanglement structure and temporal correlations, which results in stronger disturbance sensitivity than the hypergraph blockchain.

However, compared with the temporal GHZ blockchain, the proposed architecture substantially reduces quantum information requirements because classical block size no longer increases the number of required qubits.

For a blockchain containing \(B\) blocks, the temporal GHZ blockchain has \( Q_{total}^{GHZ} = O(Bm)\), hypergraph blockchain has \( Q_{total}^{HG} = O(B)\), and the proposed framework has \(Q_{total}^{Hybrid} = O(B)\).

Therefore, the scalability claim made in this work refers specifically to the efficiency in the number of qubits needed per classical block of arbitrary size - in order to be represented as a phased quantum state - rather than improvements in communication complexity or consensus throughput. Practical scalability nevertheless remains constrained by decoherence, temporal entanglement generation, recursive fusion operations, synchronization requirements, and quantum network overhead.

\subsection{Blockchain Architecture}

In the new blockchain data structure, for the classical block i, i.e. \(P_i\), phase \(\theta_{P_i}\) is calculated by the block creator much the same as in the hypergraph blockchain via a secret bijective function known only to the block creator. However, the outcome of this function is this phase \(+\) a classical \(2\)-bit string \(r_1r_2\). There are two qubits per block, initially in the ground state \(|00\rangle\). The following state is then obtained, i.e., the corresponding Bell state for the classical bit string.

\begin{equation}
|\psi\rangle = \frac{|0\,r_2\rangle + (-1)^{r_1}|1\,\bar{r_2}\rangle}{\sqrt{2}}
\end{equation}


Apply the following rotation to obtain the block if \(r_1r_2 = 00\) or \(r_1r_2 = 10\).

\begin{equation}
    |\psi_i'\rangle = S(p) |\psi\rangle = \frac{1}{\sqrt{2}}
\begin{bmatrix}
1 & 0 & 0 & 0\\
0 & 1 & 0 & 0\\
0 & 0 & 1 & 0\\
0 & 0 & 0 &e^{i\theta_p}
\end{bmatrix}
|\psi\rangle = 
\frac{|0\,r_2\rangle + e^{i\theta_{p_i}} (-1)^{r_1}|1\,\bar{r_2}\rangle}{\sqrt{2}}
\end{equation}

Otherwise, apply the following rotation.

\begin{equation}
    |\psi_i'\rangle = S(p) |\psi\rangle = \frac{1}{\sqrt{2}}
\begin{bmatrix}
1 & 0 & 0 & 0\\
0 & 1 & 0 & 0\\
0 & 0 & e^{i\theta_p} & 0\\
0 & 0 & 0 & 1
\end{bmatrix}
|\psi\rangle = 
\frac{|0\,r_2\rangle + e^{i\theta_{p_i}} (-1)^{r_1}|1\,\bar{r_2}\rangle}{\sqrt{2}}
\end{equation}


Then the \(i^{th}\) block is

\begin{equation}
|\psi_i\rangle = \frac{|0\,r_2\rangle + e^{i\theta_{p_i}} (-1)^{r_1}|1\,\bar{r_2}\rangle}{\sqrt{2}}   
\end{equation}

In the temporal case, the following state is the block of the new scheme. \citep{megidish2012resource} suggests that temporal Bell states and GHZ states can have a phase, but they emphasize that there is no need for sensitive phase accuracy: only an overlap of pulse envelopes is required for interference. Strict absolute (global) phase control is not required in the sense discussed in \citep{Megidish_2013}; however, the current work assumes that a controlled relative phase can be represented and kept stable over the verification window.

\begin{equation}
    |\psi_i\rangle^{0, \tau} = \frac{|0\,r_2\rangle^{0, \tau} + e^{i\theta_{p_i}} (-1)^{r_1}|1\,\bar{r_2}\rangle ^{0, \tau}}{\sqrt{2}}
\end{equation}

Suppose we have the first three blocks following this notation.

\begin{equation}
|\phi_1\rangle^{0, \tau} = \frac{|0\,r_{1_{2}}\rangle^{0, \tau} + e^{i\theta_{p_1}}  (-1)^{r_{1_{1}}}|1\,\bar{r_{1_{2}}}\rangle ^{0, \tau}}{\sqrt{2}}
\end{equation}

The bijective function for the genesis block is chosen such that \(r_{1_1}r_{1_2}\) is either \(00\) or \(01\) to keep the phase of the blockchain ledger below \(\pi/2\).

\begin{equation}
|\phi_2\rangle^{\tau, 2\tau} = \frac{|0\,r_{2_{2}}\rangle^{\tau, 2\tau} + e^{i\theta_{p_2}} (-1)^{r_{2_{1}}}|1\,\bar{r_{2_{2}}}\rangle ^{\tau, 2\tau}}{\sqrt{2}}
\end{equation}

\begin{equation}
|\phi_3\rangle^{2\tau, 3\tau} = \frac{|0\,r_{3_{2}}\rangle^{2\tau, 3\tau} + e^{i\theta_{p_3}} (-1)^{r_{3_{1}}}|1\,\bar{r_{3_{2}}}\rangle ^{2\tau, 3\tau}}{\sqrt{2}}
\end{equation}

These will be fused together using the fusion process in \citep{megidish2012resource}, i.e. projected recursively onto a growing temporal (time entangled) GHZ state which is the chain of blocks, using a PBS and a delay line in addition to the entangled pair generation equipment. They have suggested that strict phase control is not necessary but we have assumed it since precise phase is an integral part of our scheme\footnote{See \ref{phase} for phase control requirements in the context of the current work}. \citep{megidish2017} also implies phase in temporal GHZ states. The presence of a well-defined phase in these states depends on whether coherence can be established between different temporal modes, which we have assumed. We have also assumed temporal coherence and synchronization of the time modes \citep{cao2024}, precise control of the phase evolution of each temporal mode, minimal noise and decoherence to maintain phase stability\citep{Omran2019}, perfect entanglement \citep{Xing2023} and coherence between the temporal modes, external reference phase to define the phase, etc. We explain these assumptions in \ref{assume}. The chain of first three blocks is represented by:

\begin{equation}
\begin{aligned}
|chain\rangle^{0, \tau, \tau, 2\tau, 2\tau, 3\tau}_{r_{1_{1}}\,r_{1_{2}}\,r_{2_{1}}\,r_{2_{2}}\,r_{3_{1}}\,r_{3_{2}}} = \frac{1}{\sqrt{2}}(|0^0\,r_{1_{2}}^{\tau}\,r_{2_{1}}^{\tau}\,r_{2_{2}}^{2\tau}\,r_{3_{1}}^{2\tau}\,r_{3_{2}}^{3\tau}\rangle + \\(-1)^{r_{1_{1}}}e^{i(\theta_{P1} + \theta_{P2} + \theta_{P3})}|1^0\,\bar{r}_{1_{2}}^{\tau}\,\bar{r}_{2_{1}}^{\tau}\,\bar{r}_{2_{2}}^{2\tau}\,\bar{r}_{3_{1}}^{2\tau}\,\bar{r}_{3_{2}}^{3\tau}\rangle)
\end{aligned}
\end{equation}

Figure~\ref{fig:temporal-fusion} is the abstract schematic diagram of this fusion process for the first three blocks.

\begin{figure}[H]
\centering
\begin{tikzpicture}[
node distance=1.2cm,
box/.style={draw, rounded corners, align=center, minimum width=2.6cm, minimum height=0.9cm},
arrow/.style={->, thick}
]

\node[box] (b1) {Block 1\\ \(\theta_{P_1}, r_{1_1}r_{1_2}\)};
\node[box, below=of b1] (b2) {Block 2\\ \(\theta_{P_2}, r_{2_1}r_{2_2}\)};
\node[box, below=of b2] (b3) {Block 3\\ \(\theta_{P_3}, r_{3_1}r_{3_2}\)};

\node[box, right=2.2cm of b2] (fusion) {Recursive fusion\\ temporal projection};

\node[box, right=2.4cm of fusion] (chain) {Temporal GHZ chain\\ accumulated phase\\
\(\theta_{P_1}+\theta_{P_2}+\theta_{P_3}\)};

\draw[arrow] (b1) -- (fusion);
\draw[arrow] (b2) -- (fusion);
\draw[arrow] (b3) -- (fusion);
\draw[arrow] (fusion) -- (chain);

\end{tikzpicture}

\caption{Schematic construction of the temporal GHZ-like blockchain state. Individual phase-encoded Bell blocks are recursively fused so that the resulting chain carries the accumulated phase of the constituent blocks.}
\label{fig:temporal-fusion}
\end{figure}

The general state of the chain of blocks at \(t = n\tau\)

\begin{equation}
\begin{aligned}
|chain\rangle^{0, \tau, \tau, 2\tau,\dots (n-2)\tau, (n-1)\tau, (n-1)\tau, n\tau}_{r_{1_{1}}\,r_{1_{2}}\,r_{2_{1}}\,r_{2_{2}}\,\dots\, r_{n-1_{1}}\,r_{n-1_{2}}\,r_{n_{1}}\,r_{n_{2}}} = \frac{1}{\sqrt{2}}(|0^0\,r_{1_{2}}^{\tau}\,r_{2_{1}}^{\tau}\,r_{2_{2}}^{2\tau}\,\dots\, r_{n-1_1}^{(n-2)\tau}\, r_{n-1_2}^{(n-1)\tau}\,r_{n_1}^{(n-1)\tau}\, r_{n_2}^{n\tau}\rangle +\\ (-1)^{r_{1_{1}}}e^{i(\theta_{P1} + \theta_{P2} +\dots\theta_{P(n-1)} + \theta_{Pn})}|1^0\,\bar{r}_{1_{2}}^{\tau}\,\bar{r}_{2_{1}}^{\tau}\,\bar{r}_{2_{2}}^{2\tau}\,\dots\, \bar{r}_{n-1_{1}}^{(n-2)\tau}\,\bar{r}_{n-1_{2}}^{(n-1)\tau}\,\bar{r}_{n_{1}}^{(n-1)\tau}\,\bar{r}_{n_{2}}^{n\tau}\rangle)
\end{aligned}
\end{equation}

where \(r_{1_1}\,r_{1_2}\) is either \(00\) or \(01\).

There are constraints on the phase similar to the hypergraph blockchain as part of the consensus. Consensus will be discussed later. The first is \(\sum_{i = 1}^\infty \theta_{P_i} < \frac{\pi}{2}\), following \citep{shreya2020}, in order to create a basis for consensus.
\begin{equation}
\theta_{P_i} = \ \frac{\theta_{P_1}}{n^{(i-1)}} 
\end{equation}

where \( n \in \mathbb{N} \setminus \{1\} \). \(\theta_{P_1}\) is shared with the entire network by the creator of the genesis block along with \(r_{1_1}r_{1_2}\). \(\sum_{i = 1}^\infty \theta_{P_i} = \sum_{m = 1}^\infty \frac{\theta_{P_1}}{(n - 1)^m} = \frac{n}{n - 1}\theta_{P_1}\). The upper bound is \(\pi/2\). Therefore when \(n = 2\), it needs to be \(\theta_{P_1} < \pi/4\). A suitable \(\theta_1\) needs to be chosen with the chosen \(n\).

\begin{table}[H]
\caption{Protocol summary for the proposed hybrid quantum blockchain framework.}
\label{tab:protocol-summary}

\begin{tabularx}{\textwidth}{|p{0.22\textwidth}|X|}
\hline

\textbf{Step} & \textbf{Operation} \\

\hline

1. Initialization &
The genesis block creator selects \(\theta_{P_1}\), \(n\in\mathbb{N}\setminus\{1\}\), and an initial classical string \(r_{1_1}r_{1_2}\), subject to \(\theta_{P_i}=\theta_{P_1}/n^{i-1}\) and \(\sum_i\theta_{P_i}<\pi/2\). \\

\hline

2. Classical-to-phase mapping &
For a classical block \(P_i\), the block creator applies a secret bijective mapping that outputs a phase \(\theta_{P_i}\) and a two-bit string \(r_{i_1}\,r_{i_2}\). \\

\hline

3. Bell-state block preparation &
The block creator prepares and distributes the two-qubit state

\[
|\psi_i\rangle=
\frac{|0\,r_{i_2}\rangle+
e^{i\theta_{P_i}}(-1)^{r_{i_1}}
|1\,\bar r_{i_2}\rangle}{\sqrt2}
\]

\\\\

\hline

4. State distribution &
The proposer distributes verification copies of the candidate quantum block to validators through quantum communication channels and distributes the corresponding classical string \(r_{i_1}r_{i_2}\) and metadata through QKD-secured classical channels. \\

\hline

5. Verification measurement &
Each validator reconstructs the expected phase \(\theta_{P_i}\) from the predetermined phase series, forms the corresponding phase-dependent Bell basis, and measures received verification copies of the received state. \\

\hline

6. Acceptance rule &
A validator outputs \(ACCEPT\) if the observed verification probability exceeds threshold \(P_{th}\). In the ideal noiseless case:

\[
P_{th}=1
\]

In realistic settings:

\[
P_{th}<1
\]

must be selected according to experimental noise levels. \\

\hline

7. Consensus and finalization &
Validators broadcast \(ACCEPT\) or \(REJECT\). The block is committed when at least \(2f+1\) validators accept, assuming \(N\geq3f+1\). The remaining valid copy is appended to the validator's local quantum ledger. \\

\hline

8. Temporal linking &
Successive phase-encoded Bell states are recursively fused, following the temporal-entanglement construction of \citep{megidish2012resource}, into a growing temporal GHZ-like chain whose accumulated phase contains \(\sum_i\theta_{P_i}\). \\
\hline

\end{tabularx}

\end{table}

\begin{figure}[H]
\centering

\begin{tikzpicture}[
node distance=1.4cm,
box/.style={draw, rounded corners, align=center, minimum width=3.0cm, minimum height=1cm},
arrow/.style={->, thick}
]

\node[box] (data) {Classical block\\ \(P_i\)};
\node[box, right=of data] (phase) {Phase mapping\\ \(\theta_{P_i}, r_{i_1}r_{i_2}\)};
\node[box, right=of phase] (bell) {Phase-encoded\\ Bell block};
\node[box, below=of bell] (verify) {Validator measurement\\ phase-dependent basis};
\node[box, left=of verify] (consensus) {Consensus\\ \(2f+1\) ACCEPT};
\node[box, left=of consensus] (fusion) {Recursive fusion\\ temporal GHZ chain};

\draw[arrow] (data) -- (phase);
\draw[arrow] (phase) -- (bell);
\draw[arrow] (bell) -- (verify);
\draw[arrow] (consensus) -- (fusion);
\draw[arrow] (verify) -- (consensus);

\end{tikzpicture}

\caption{Schematic flow of the proposed hybrid quantum blockchain framework. Classical block data are mapped into phase values and two-bit strings, embedded into phase-encoded Bell states, recursively fused into a temporal GHZ-like blockchain state, verified using phase-dependent measurements, and finalized by validator consensus.}
\label{fig:protocol-flow}
\end{figure}

Table \ref{tab:protocol-summary} and Figure \ref{fig:protocol-flow} outline the protocol flow of the proposed framework. Similar to the GHZ blockchain \citep{rajan2019} and the hypergraph blockchain \citep{shreya2020}, it must be understood that the blockchain architecture proposed here is a conceptual framework, and practical implications have not yet been fully considered. This work
should be viewed as a foundation for future research on the refinement, extension, and correction of this theoretical model or the practical realization of this blockchain framework. When algorithms Shor \citep{shor1994} and Grover \citep{needle1997} were conceptualized, they were entirely theoretical for quantum computers were far from realization; this framework must be viewed from the same perspective.

\subsection{Quantum Network}

Similar to both \citep{shreya2020} and \citep{rajan2019}, the proposed framework assumes both classical and quantum communication layers \citep{Simon2017}. The role of each layer is distinct and we will separate these roles as follows.

\paragraph{Quantum Communication Layer}

Quantum channels are required for distribution of quantum block states, distribution of quantum public-key states used by quantum signature protocols, and execution of quantum operations required by consensus and verification. These channels are not assumed perfectly secure by themselves. Rather, security derives from the properties of the protocols operating on top of them.

\paragraph{QKD}

Quantum key distribution is assumed only for generating symmetric keys used to encrypt classical communications including validator communication, consensus messages, transaction transmission, and distribution of classical metadata required for verification. Therefore QKD provides confidentiality and secure symmetric key establishment. QKD alone does not provide authentication, transaction authorization, or non-repudiation.

\paragraph{Quantum Signature}

The quantum digital signature protocol of \citep{gottesman2001quantumdigitalsignatures} is assumed for transaction signing, node authentication, message authentication, and non-repudiation. 

Therefore, signatures and QKD provide different security functions. QKD protects communication confidentiality, whereas signatures provide authentication and authorization. The motivation for using quantum signatures rather than classical signatures is that classical signature security ultimately relies on computational assumptions, whereas quantum signature constructions aim to provide stronger security assumptions against adversaries possessing large-scale quantum computation. That is, quantum signatures in the literature attempt to provide information-theoretic guarantees rather than computational security guarantees.

\paragraph{Relationship to Classical Blockchains}

Classical blockchains already provide integrity through cryptographic hashes, signatures, and consensus. The proposed framework does not claim to replace all classical security mechanisms.

Instead, the quantum blockchain primitives in place of classical counterparts primarily introduce disturbance detectability through quantum state verification and resistance against blockchain mutation through physically encoded quantum block states. Both of these aim to overcome the theoretical possibility of a classical blockchain being mutated by a computationally powerful adversary. That is, a powerful adversary can, in theory, mutate a block and forge subsequent classical hashes in a sufficiently short amount of time. The purpose of a quantum consensus protocol is to facilitate the interaction of quantum primitives used in the framework, encompassing the distribution and verification of quantum states. A vote-based classical consensus protocol\cite{xu2023, deng2022} could be tailored to be used in this framework with modification for quantum state distribution and verification.

Therefore, the intended advantage is not merely stronger computation but the introduction of physical mechanisms for disturbance detection quantum-native authentication.

\paragraph{Public-Key Distribution and Decentralization}

The original Gottesman construction discusses trusted quantum public-key distribution. However, relying on a centralized trusted authority conflicts with blockchain decentralization. Therefore this work does not require a centralized authority. Instead, quantum public-key distribution is assumed achievable through decentralized quantum network infrastructure capable of distributing authenticated quantum states between validators. The practical realization of such decentralized quantum public-key infrastructure remains an open problem and is treated as an assumption of the present conceptual framework.

Space-time and satellite effects \citep{bruschi2014}, quantum network architectures \citep{manssor2024,diamado2021}, and entangled-network capacity constraints \citep{jiang2024} remain important research directions for realizing such infrastructure.

\subsection{Consensus}

Consensus, at a high level, is similar to the hypergraph blockchain. To validate the phase, the block creator distributes a large number of copies of the state of the \(m\)th quantum block to the entire network, i.e. the state 
\(\frac{|0\,r_{m_{1}}\rangle + (-1)^{r_{m_1}}e^{i\theta_m}|1\,\bar{r}_{m_{1}}\rangle}{\sqrt{2}} \),
is distributed to each network node without violating the no-cloning theorem, and each node can use all except one to check the validity and the other to append the next block to their local chain. The classical string \(r_{m_1}r_{m_2}\) (to be securely stored) is distributed by classical channels secured via QKD. This state distribution is done via the mutual quantum channels. At this stage of the design, following \citep{rajan2019}, we assumed that newly generated blocks are spatial GHZ states since entangling these qubits in time at this stage of the design process is unnecessary and is left for future work. 
The block producer is chosen randomly, utilizing a low-level algorithm based on Quantum Random Number Generation (QRNG) \citep{Mannalatha_2023}.

\begin{equation}
|\phi'_m\rangle = \frac{|0\,r_{m_{1}}\rangle + (-1)^{r_{m_1}}e^{i\theta_m}|1\,\bar{r}_{m_{1}}\rangle}{\sqrt{2}}
\end{equation}

\paragraph{Verification basis}

To check the validity of this block, each receiving node measures the quantum block \((\frac{|0\,r_{m_{1}}\rangle + (-1)^{r_{m_1}}e^{i\theta_m}|1\,\bar{r}_{m_{1}}\rangle}{\sqrt{2}})\) in the basis \(\frac{|0\,r_{m_{1}}\rangle \pm (-1)^{r_{m_1}}e^{i\theta_{m_{predetermined}}}|1\,\bar{r}_{m_{1}}\rangle}{\sqrt{2}}\) \(+\) the other two corresponding states that together with these two states form an orthonormal basis: the Bell basis with an added phase. If the outcome is not deterministic with the probability of \(1\), the phase is considered invalid, and the block creator is deemed untrustworthy by the validating node. Then the validating node shares its judgment with the network along with measurement results: the phase is invalid, different nodes have received different phases/states, or different nodes have received different classical strings. With the knowledge of the bit string and the predetermined phase based on \(m\), \(\theta_1\), and ratio \(n\), a node can reconstruct the measurement basis for a number of measurements to calculate this probability. This effectively reduces block validation to a quantum state verification problem under a known phase constraint. In this basis, \(\theta_m\__{pre} = \frac{\theta_1}{n^{(m-1)}}\); \(\theta_{P_1}\) is shared openly by the creator of the genesis block. If valid, the last copy can be appended to the local blockchain.

The proposed blockchain operates as a permissioned vote-based consensus protocol consisting of \(N\) validator nodes.

After local verification, each validator broadcasts \( ACCEPT \) or \(REJECT\), together with:

\begin{itemize}

\item local verification outcome

\item phase consistency result

\item classical string consistency result

\end{itemize}

A block becomes committed only when at least \(2f+1\) validators broadcast \(ACCEPT\), where \( N \geq 3f+1\) and \(f\) denotes the maximum tolerated Byzantine validators. Under this assumption, the protocol tolerates \(f < \frac{N}{3}\) Byzantine validators.

The protocol therefore proceeds through four phases:

\begin{enumerate}

\item QRNG-based proposer selection

\item Quantum block distribution

\item Local verification

\item Vote collection and finalization

\end{enumerate}

Message complexity is \(O(N^2)\)

since validators broadcast decisions to all validators.

\paragraph{Safety Argument}

Suppose fewer than \(f\) validators behave Byzantine. Since commitment requires \(2f+1\) votes, two conflicting blocks cannot simultaneously obtain quorum. Therefore conflicting finalized ledgers cannot occur under the stated assumptions. A complete proof of safety, liveness, and Byzantine fault tolerance remains future work. The temporal GHZ state used by the framework enables nonclassical correlations between records generated and exist at different temporal instances while consensus itself remains dependent on validator agreement and state verification.

\subsection{Threat Model and Security Analysis} \label{threat}

\subsubsection{Adversarial Model}

The proposed framework assumes a partially synchronous permissioned network consisting of \(N\) validator nodes connected through authenticated classical channels and quantum communication links.

We assume adversaries may possess the following capabilities:

\begin{itemize}

\item Intercept quantum transmissions

\item Perform local measurements on received quantum states

\item Distribute inconsistent block states to different validators

\item Control up to \(f\) validator nodes

\item Delay classical communications

\end{itemize}

The adversary is assumed incapable of:

\begin{itemize}

\item Compromising quantum key distribution

\item Cloning arbitrary quantum states due to the no-cloning theorem

\item Simultaneously controlling more than the tolerated number of validators

\end{itemize}

Security goals of the protocol are therefore restricted to:

\begin{itemize}

\item Detection of unauthorized measurements

\item Detection of inconsistent state distribution

\item Agreement among honest validators

\item Preservation of ledger consistency

\end{itemize}

The protocol does not claim confidentiality of blockchain contents nor unconditional security against arbitrary quantum adversaries.

\subsubsection{Disturbance Detectability}

Let the transmitted block state be

\begin{equation}
   |\psi_i\rangle=\frac{|0\,r_2\rangle+e^{i\theta_{P_i}}(-1)^{r_1}|1\,\bar r_2\rangle}{\sqrt 2}
\end{equation}

Validators measure received states in the expected orthonormal basis.

Suppose an adversary performs an unauthorized projective measurement \(M\).

The original pure state density matrix \(\rho=|\psi_i\rangle\langle\psi_i|\) becomes \( \rho'= \sum_k P_k \rho P_k \), where \(P_k\) are projectors associated with measurement \(M\). Because measurement removes phase coherence terms, \(\rho' \neq \rho \), and therefore, the probability of obtaining deterministic outcomes under the expected verification basis decreases. Consequently, unauthorized measurements become detectable through verification statistics. This property provides disturbance detectability rather than information-theoretic security.

\subsubsection{Detection-Probability Example}

We consider the ideal encoded block state

\begin{equation}
|\psi(\theta)\rangle =
\frac{|0r_2\rangle+
e^{i\theta}(-1)^{r_1}|1\bar r_2\rangle}{\sqrt{2}} .
\end{equation}

Validators verify the state by projecting onto the expected state
\(|\psi(\theta, r_1r_2)\rangle\). If an adversary introduces a phase perturbation \(\delta\), the received state becomes

\begin{equation}
|\psi(\theta+\delta, r_1r_2)\rangle =
\frac{|0r_2\rangle+
e^{i(\theta+\delta)}(-1)^{r_1}|1\bar r_2\rangle}{\sqrt{2}} .
\end{equation}

The probability that this perturbed state passes one ideal verification test is

\begin{equation}
P_{\mathrm{accept}}
=
|\langle \psi(\theta, r_1r_2)|\psi(\theta+\delta), r_1r_2\rangle|^2
=
\cos^2\left(\frac{\delta}{2}\right).
\end{equation}

Therefore, the ideal single-copy detection probability is

\begin{equation}
P_{\mathrm{detect}}
=
1-
P_{\mathrm{accept}}
=
\sin^2\left(\frac{\delta}{2}\right).
\end{equation}

If \(C\) independent verification copies are tested, the probability that at least one test detects the perturbation is

\begin{equation}
P_{\mathrm{detect}}^{(C)}
=
1-
\left[
\cos^2\left(\frac{\delta}{2}\right)
\right]^C .
\end{equation}

As a second simple case, we imagine an adversary performs a projective measurement in the computational basis. The post-measurement state becomes an incoherent mixture, and its acceptance probability under the original Bell-basis verification is

\begin{equation}
P_{\mathrm{accept}}^{\mathrm{meas}}
=
\frac{1}{2}.
\end{equation}

Thus, with \(C\) independent verification copies, the corresponding ideal detection probability is

\begin{equation}
P_{\mathrm{detect}}^{\mathrm{meas},C}
=
1-
\left(\frac{1}{2}\right)^C .
\end{equation}

These quantitiers are idealized and assume noiseless preparation, transmission, and measurement. Under realistic noise, the acceptance rule must be replaced by a threshold rule, as discussed below.

\subsubsection{Noise and Verification Reliability}

The verification procedure assumes that deviations from expected measurement statistics may arise from both adversarial tampering and experimental imperfections. Suppose the ideal verification probability is \(P_{ideal}=1\). Under realistic conditions with no adversarial presence in the network, the observable verification probability \((P_{obs} < 1)\) may occur because of phase noise, imperfect state preparation, decoherence, detector errors, photon loss, and imperfect Bell measurements or fusion operations.

Consequently, failed verification cannot always be interpreted as evidence of malicious tampering. Instead, practical implementations would require acceptance thresholds \(P_{obs}\geq P_{th}\), where \(P_{th}<1\), and \(P_{th}\) is determined by experimentally achievable noise levels.

Therefore, the disturbance detectability discussed in this work should be interpreted as an idealized - zero operational noise - conceptual property rather than an experimentally characterized security guarantee under realistic noise.

\subsubsection{Security Limitations}

The proposed framework presently provides only conceptual guarantees regarding disturbance detectability. The security properties that are not formally proven in this work include, Byzantine fault tolerance under arbitrary adversaries, liveness guarantees, resistance against coordinated validator collusion, formal composable security, and security under imperfect quantum hardware.
Therefore, the current framework should be interpreted as a conceptual architecture rather than a complete secure blockchain protocol.

\subsection{Disturbance Detectability and State Integrity} 
All nodes share the measurement outcomes, judgement on the block creator's honesty, and \(r_{i_1}r_{i_2}\) they received from the block creator with all other nodes. Then based on other nodes' judgements and information, a node shares their final verdict on the admissibility of the block. These validating nodes can compare the classical 2-bit strings \(r_{1}r_{2}\) shared by other nodes and determine if there are signs of the block creator having distributed different bit strings to different nodes. If such a receiving node is dishonest, they could either report the block creator as dishonest when the new block is valid or report honest when the new block's phase is invalid. However, as long as more than half of the nodes are honest, their judgment on the admissibility of the block will be accepted; this is attributable to the fact that all honest nodes will come to the same conclusion about the admissibility. Nodes whose verdicts disagree with the accepted consensus outcome are treated as potentially faulty or adversarial.  

In the proposed quantum blockchain framework, the traditional time-stamped blocks and hash functions connecting them are replaced by a temporal GHZ state with a growing phase, utilizing entanglement over time. Hence, the sensitivity to tampering is increased. If a single block/qubit is measured, the entire local copy of the blockchain is compromised due to the entanglement, whereas, in a classical blockchain, only the blocks following the tampered one are affected (hash pointers), leaving the system susceptible to vulnerabilities. In classical blockchains, it is often claimed that the further back a block is in the chain, the more secure it becomes. This is because tampering with a block earlier in the chain invalidates more data, thus strengthening security (immutability). The following section elaborates on this phenomenon of GHZ entanglement

\subsection{Theoretical Tamper Sensitivity of the GHZ Blockchain in Comparison to the Quantum Hypergraph Blockchain\texorpdfstring{\citep{shreya2020}}{[Shreya et al., 2020]}}

It is important to note that weighted hypergraph states \cite{shreya2020} and GHZ states possess different multipartite entanglement structures and might not be equal in terms of entanglement correlation strength. We elaborate on that in this section. A phased GHZ state
\begin{equation}
|{\rm GHZ}_{\phi}\rangle = \frac{1}{\sqrt{2}}\left(|0\ldots 0\rangle +
e^{i\phi}|1\ldots 1\rangle\right)
\end{equation}
remains maximally entangled for any value of the phase $\phi$, and is extremely fragile under local measurements: a projective measurement on any single qubit collapses the entire state and destroys the entire multipartite entangled state.

\paragraph{The Non-Maximal Entanglement of Weighted Hypergraph States}
According to \citep{yamazaki2025}, a weighted graph state (WGS), a subclass of Weighted Quantum Hypergraph States (WQHS), is generated when controlled-phase (CP) gates with an angle $\phi \neq \pi$ are used, instead of the maximally entangling Controlled-Z ($\text{CZ}$) gates, where $\phi = \pi$. This implies that the resulting state is generally not maximally entangled. This paper, "Measurement-based quantum computation on weighted graph states with arbitrarily small weight,"\citep{yamazaki2025} explicitly addresses the nature of the entangling gates and the resulting state:

\begin{quote}
    "Weighted graph states are a natural generalization of graph states, which are generated by applying controlled-phase gates, instead of controlled-Z gates, to a separable state. ... One such experimental constraint is that only controlled-phase ($\text{CP}$) gates with angle $\phi$ are available, instead of controlled-Z ($\text{CZ}$) gates ($\phi = \pi$). Then, a generated state becomes a so-called \textbf{weighted graph state} (WGS), instead of a graph state." \citep{yamazaki2025}
\end{quote}

The authors further classify the entanglement inherent to these states, stating that WGS are the first example of a universal resource created this way:

\begin{quote}
    "To our knowledge, this is the first example of universal resources prepared with only \textbf{non-maximally entangling gates} and has potential applications to weakly interacting systems, such as photonic systems." \citep{yamazaki2025}
\end{quote}

This use of \textbf{non-maximally entangling gates} to prepare the WGS (and, more generally, WQHS with variable weights) suggests that the resulting states may not always be maximally entangled.

In contrast to GHZ, the weighted hypergraph states employed in the hypergraph blockchain construction \citep{shreya2020} have a significantly more distributed and non-uniform entanglement structure. As stated above, these states include amplitudes for all computational basis states,
\begin{equation}
|\psi\rangle = \frac{1}{\sqrt{2^{N}}}\sum_{x\in\{0,1\}^{N}}
e^{i\pi f(x)}|x\rangle
\end{equation}
with phases determined by the pattern and weights of multiple $k$-hyperedges. Because the entanglement in such states is typically non-maximal and is spread across many local correlations, their response to local measurements is less catastrophic than that of a GHZ state. In general, measuring a single qubit of a weighted hypergraph state \textbf{theoretically} does not annihilate all global correlations, and the post-measurement state may still retain nontrivial entanglement among the remaining qubits.

\paragraph{Measurement and Unitary Tampering Model}

Consider an adversary capable of performing unauthorized local operations on accessible qubits. Tampering is defined as any operation that modifies the verification statistics observed by honest validators. We consider two classes of adversarial operations:

\begin{enumerate}

\item Local projective measurements

\item Local unitary transformations

\end{enumerate}

Let \(\rho\) denote the density matrix representing the legitimate blockchain state. Suppose the adversary acts on qubit \(i\) using operator \(T_i\) where \(T_i\in \{M_i,U_i\}\), where \(M_i\) is a local projective measurement operator and \(U_i\) is a local unitary operator.

The resulting state becomes \(\rho' =  T_i\,\rho\, T_i^\dagger\) for unitary operation tampering, or \(\rho' =  \sum_k P_k\, \rho\, P_k\) for projective measurements, where \(P_k\) are projectors associated with measurement \(M_i\). For phased GHZ states:

\begin{equation}
    |{\rm GHZ}_{\phi}\rangle=
\frac{
|0\dots 0\rangle+
e^{i\phi}|1\dots 1\rangle
}{
\sqrt2
}
\end{equation}

,the coherence between the two branches depends globally on all participating qubits. Therefore local measurements remove coherence terms, while local unitary operations generally modify global phase relationships required for verification. Consequently, it holds that \(\rho'_{GHZ}\neq \rho_{GHZ}\), and verification statistics dependent on multipartite coherence change immediately.

Weighted hypergraph states generally possess distributed correlations:

\begin{equation}
|\psi\rangle=
\frac1{\sqrt{2^N}}
\sum_x
e^{i\pi f(x)}
|x\rangle
\end{equation}

where \(N\) is the number of qubits and \(f(x)\) is the hyperedge-dependent phase function. Because correlations are distributed across multiple amplitudes, local operations need not necessarily destroy all nonclassical correlations simultaneously.

Under this tampering model, GHZ states therefore exhibit stronger immediate disturbance under local operations because coherence depends globally on all participating qubits. This comparison is qualitative rather than quantitative. We do not claim a general security ordering between GHZ states and weighted hypergraph states, and we do not establish superiority under a specific multipartite entanglement measure. Rather, the observation is limited to disturbance detectability under the local tampering model considered above. A rigorous quantitative comparison using entanglement measures and attack-specific security metrics remains a direction for future work.

It is worth emphasizing that although weighted hypergraph states belong to the class of LME states, this property does not in general imply maximal multipartite entanglement. In the sense used in this work, LME refers to the ability of a state to generate maximal entanglement between the system and a set of local ancillas via suitable local unitary operations. This is an operational notion that concerns the entangling power of the state under local control, rather than the intrinsic amount of global entanglement present in the multipartite system itself. In particular, the degree of genuine multipartite entanglement depends sensitively on the choice of hyperedge weights and the resulting phase structure. For generic weight assignments, the reduced density matrices of subsystems need not be maximally mixed, and standard entanglement measures need not attain their maximal values. Consequently, while the LME property ensures strong nonclassical correlations and local controllability, it does not guarantee that the corresponding weighted hypergraph states are maximally entangled in the multipartite sense.

Hence, under the measurement-based tampering model considered in this work, GHZ-based constructions may provide stronger immediate disturbance detectability under local measurements than weighted hypergraph constructions. This comparison should not be interpreted as a complete security ordering between GHZ states and weighted hypergraph states. Rather, the comparison establishes that the proposed framework inherits a disturbance mechanism associated with globally distributed multipartite coherence that is not obviously present to the same extent in weighted hypergraph constructions.

\subsection{The Spatial GHZ Blockchain Ledger Structure and Disturbance Detectability: Without the Temporal Aspect}

The data is encoded into the phase of the GHZ state and the basis states themselves. Therefore, for an adversary to measure the phase, they will need to keep measuring the state in random basis as long as the length of the chain, \(\theta_{P_1}\), classical 2-bit strings, or \(n\) remain secrets. This randomness includes GHZ basis states computational basis states as well. Even if the chain is constructed by the adversary using those parameters (even if the adversary obtains bit strings and the phases of blocks), without the knowledge of the bijective functions, it is not possible to obtain the data. Even if the data is obtained, i.e., the bijective functions are discovered, it is still not possible to mutate the data with that same bijective function intact. That is, it is impossible to change the phase of the \(i^{th}\) block, i.e., \(\theta_{Pi}\), and the classical string \(r_{i_1}r_{i_2}\), since this would result in a different phase other than that which is predefined for the GHZ state. And it would also result in different basis states that do not represent the classical bit strings \(r_{i_1}r_{i_2}\). If the length of the chain is \(m\), i.e. \(2m\) qubits, the projection measurement basis to check validity will be

\begin{equation}
\frac{1}{\sqrt{2}}(|0\,r_{1_{2}}\,r_{2_{1}}\,r_{2_{2}}\,\dots\,r_{n-1_1}\,r_{n-1_2}\,r_{n_1}\,r_{n_2}\rangle \pm (-1)^{r_{1_{1}}}e^{i\theta_{expected}}|1\,\bar{r}_{1_{2}}\,\bar{r}_{2_{1}}\,\bar{r}_{2_{2}}\,\dots\,\bar{r}_{n-1_{1}}\,\bar{r}_{n-1_{2}}\,\bar{r}_{n_{1}}\,\bar{r}_{n_{2}}\rangle)
\end{equation}

with the rest of the basis set: the other \(2^{2n} - 2\) states that make an orthonormal basis (GHZ basis with an added phase) with these 2 states, where in the projection \(\{\Omega, I - \Omega\}\), \(\Omega\) is the projection operator for the \(+\) state \textbf{out of the two states above}.

Also, in the above, \(\theta_{expected} = \sum_{i=1}^m\theta_m = \sum_{i=1}^m \frac{\theta_1}{n^{(m-1)}}\), and \(\theta_{P_1}\), i.e., \(\theta_{1}\), are shared with the entire network by the creator of the genesis block. Since the phase is the phase of the whole compound GHZ state, when changing the phase of the \(i^{th}\) block, it is suggested that \(\theta_{Pi}\) is changed, and therefore the phase of the entire GHZ state is changed, not the phase of individual pairs of qubits or single qubits.

Measurement in the correct basis will reveal whether the phase and GHZ basis states are valid, i.e. the predefined phase and locally stored classical 2-bit strings. The outcome must be deterministic for the basis state that the chain of blocks should be in with probability no less than \(1\)\footnote{In the ideal noiseless model, acceptance is deterministic; practical systems would require thresholded acceptance probabilities.}. Any other outcome reveals that the chain is invalid for the current length and the owner of that local copy can reconstruct the blockchain ledger with the use of the length of the chain, \(\theta_{P_1}\), 2-bit classical strings received from block creators, and \(n\). If there is any suspicion of mutation, reconstruction can be done at any point. Depending on the exact application built on the blockchain architecture, the secret bijective functions would need to be shared with the network; certain functions are disclosed to certain nodes. 

In the temporal case, only the last remaining qubit, which holds the entire aggregated phase of the entire chain, can be modified in its phase or GHZ basis states: the above arguments and theory apply accordingly. For instance, the basis will account for the last qubit with the phase being the same as above.

With just a spatial GHZ state, measurement correlations are stronger than any classical blockchain. The chain being a maximally entangled state, any measurement on any of the qubits will lead to a collapse of the entire local copy. In this scenario, if an attacker alters any photon, the entire local copy of the blockchain is immediately invalidated (collapses), offering a clear advantage over classical systems, where only blocks after the tampered one are affected. With the temporal aspect added to the GHZ chain, past photons no longer exist, preventing any attempt to access or tamper with them. An attacker can only alter the final remaining photon, which would immediately invalidate the entire state. 

The temporal GHZ state in this blockchain framework involves an entanglement between photons that do not share simultaneous coexistence \citep{Megidish_2013}, yet they share nonclassical measurement correlations. It can be interpreted as a way to link records in the current block to records not just representative \textit{of} the past but only existed \textit{in} the past.

A case by case analysis of security is still necessary and is left for future.

\subsection{Local Unitary Operations on the chain of blocks}

The spatially entangled blockchain data structure is accounted for in this case. The aim of behind applying these transformations the further solidification of difficulty of tampering. This obscures the actual state of the chain from attackers, making it impossible to mutate the chain, i.e., mutate the phase. Also, this aims to make it impossible to extract data even with the knowledge of the initial phase, the length \(m\) of the \(2m\) qubit chain, and the bijective functions. That is, the adversary would not be able to obtain the chain by reverse transformations or construct the chain and therefore the phase of the blockchain. Peers can apply local unitary operations to the subsystems of the quantum blockchain data structure, which is a compound quantum state. In the case of a local unitary operation to a subsystem, the operation would always be reversible since unitary operations are reversible, but depending on the actual transformation applied, the phase, after the reversal, might not exactly be the same as the original phase. However, a global unitary operation such as a phase shift on the whole state would be reversible to the same initial blockchain. However, if the inverse transformation cannot restore the original state, with the knowledge of \(n\), chain length \(m\), all the \(2\)-bit strings, and \(\theta_{P1}\), an honest peer would be able to reconstruct the blockchain.

In the temporal case, since only one qubit with the cumulative phase exists, any rotation or unitary transformation is possible and reversible.

\section{Comparison of Quantum Blockchain Architectures}

To contextualize the proposed framework within the landscape of quantum blockchain research, we present a systematic comparison across the two frameworks that inspired our work and the proposed work. Table \ref{tab:comparison} summarizes the essential characteristics of the temporal GHZ blockchain \citep{rajan2019}, the quantum hypergraph blockchain \citep{shreya2020}, and the proposed hybrid architecture.
\begin{table}[H]
\caption{Comparison of quantum blockchain architectures.\label{tab:comparison}}
\begin{adjustwidth}{-\extralength}{-\extralength}
\centering
\begin{tabularx}{\dimexpr\textwidth+2\extralength\relax}{p{2.5cm}CCC}
\toprule
\textbf{Property} & \textbf{Temporal GHZ \citep{rajan2019}} & \textbf{Quantum Hypergraph \citep{shreya2020}} & \textbf{Proposed Framework}\\
\midrule

Encoding Efficiency &
2 classical bits $\rightarrow$ 2 qubits (1:1 ratio) &
Arbitrary-sized classical block $\rightarrow$ 1 qubit &
Arbitrary-sized classical block $\rightarrow$ 2 qubits\\
\midrule

Quantum State Structure &
Temporal GHZ state: $\frac{1}{\sqrt{2}}(|0\cdots0\rangle + (-1)^{r_1}|1\cdots\overline{1}\rangle)$ &
Weighted hypergraph state: $\frac{1}{\sqrt{2^N}}\sum_{x} e^{i\pi f(x)}|x\rangle$ &
Phased temporal GHZ: $\frac{1}{\sqrt{2}}(|0\,r_2\cdots\rangle + e^{i\theta}(-1)^{r_1}|1\,\bar{r}_2\cdots\rangle)$\\
\midrule

Entanglement Type &
Maximal multipartite (temporal GHZ) &
Typically Non-maximal, distributed (weighted LME states) &
Maximal multipartite (phased temporal GHZ)\\
\midrule

Temporal Entanglement &
Yes (past blocks cease to exist) &
No (all qubits coexist spatially) &
Yes (past blocks cease to exist)\\
\midrule

Disturbance Sensitivity &
Maximal: any measurement collapses entire chain; past blocks inaccessible &
Reduced: distributed entanglement may survive partial measurements &
Maximal: measurement collapses chain; past blocks inaccessible; phase adds verification layer\\
\midrule

Data Encoding Mechanism &
Computational basis states only $(r_1, r_2)$ &
Phase encoding via bijective function $\theta_P=f(P)$ &
Hybrid: basis states $(r_1, r_2)$ + phase $\theta_P=f(P)$\\
\midrule

Phase Constraints &
None &
$\sum_{i=1}^{\infty}\theta_{P_i}<\frac{\pi}{2}$, with $\theta_{P_i}=\frac{\theta_{P_1}}{n^{i-1}}$ &
$\sum_{i=1}^{\infty}\theta_{P_i}<\frac{\pi}{2}$, with $\theta_{P_i}=\frac{\theta_{P_1}}{n^{i-1}}$\\
\midrule

Encoding Efficiency (qubit cost) &
$O(m)$ qubits for $m$-bit classical data &
$O(1)$ qubits per block (1 qubit/block) &
$O(1)$ qubits per block (2 qubits/block)\\
\midrule

Consensus Validation &
Measurement in computational/GHZ basis; deterministic outcome required &
Measurement in phase-shifted basis $\frac{|0\rangle \pm e^{i\theta_{\text{expected}}}|1\rangle}{\sqrt{2}}$; probabilistic verification &
Measurement in phased GHZ basis with basis states $\{|0\,r_2\rangle,|1\,\bar{r}_2\rangle\}$; combined validation\\
\midrule

Block Distribution &
Bell pairs distributed to network; fusion at nodes &
Single qubit distributed (spatial GHZ copies for validation) &
Phased Bell pairs distributed; spatial phased GHZ copies for validation\\
\midrule

Secret Information &
2-bit strings $r_1r_2$ per block &
Bijective function $f:P\rightarrow\theta_P$ per block creator &
Both: 2-bit strings $r_1r_2$ + bijective function $f$\\
\midrule

Attack Surface &
Only last photon exists; measurement immediately detectable via collapse &
All qubits exist; phase can theoretically be probed via repeated measurements &
Only last photon exists; both phase and basis must be validated;\\
\midrule

Reconstruction Capability &
Requires stored $r_1r_2$ strings, $\theta_{P_1}$, $n$, chain length &
Requires stored phases, bijective functions &
Requires stored $r_1r_2$ strings, $\theta_{P_1}$, $n$, bijective functions, chain length\\
\midrule

Disturbance (Tamper) Detectibility Basis &
Temporal inaccessibility + measurement collapse &
Phase encoding obfuscation + entanglement &
Temporal inaccessibility + measurement collapse + phase encoding + basis state verification\\
\midrule

Experimental Feasibility (Current) &
Temporal Bell/GHZ states demonstrated \cite{megidish2012resource}; phase control not addressed &
Spatial hypergraph states realizable; phase-encoding demonstrated &
Requires temporal entanglement + precise phase control + recursive fusion; not yet demonstrated\\
\midrule

Practical Assumptions &
Perfect temporal entanglement, negligible decoherence, ideal fusion &
Perfect spatial entanglement, ideal phase gates, controlled CP operations &
Perfect temporal entanglement, strict phase control, temporal coherence, synchronization, negligible decoherence\\

\bottomrule
\end{tabularx}
\end{adjustwidth}
\end{table}
\section{Key Assumptions}\label{assume}

\subsection{Phase Definition and Phase Control} \label{phase}

The proposed framework does not require absolute strict phase control in the sense of fixing an externally meaningful global phase. This is important because \citep{megidish2012resource} explicitly indicates that their multiphoton entanglement source does not rely on sensitive phase stabilization; instead, interference requires indistinguishability and temporal overlap of the relevant photon wave packets.

Therefore, the phase parameter \(\theta_{P_i}\) used in this work should be interpreted as a controllable relative phase used for encoding and verification, not as an uncontrolled absolute optical phase. The practical requirement is that the relative phase between the two components of the encoded Bell or GHZ-like state remains stable over the verification window:

\begin{equation}
|\Delta \theta_{P_i}| < \epsilon_{\theta}
\end{equation}

where \(\epsilon_{\theta}\) denotes the tolerated phase error determined by the verification basis.

If phase drift exceeds this tolerance, honest states may be rejected or invalid states may pass verification. Thus, phase stability is a feasibility constraint rather than an assumed solved problem. Recent work on time-bin and temporal photonic systems shows that phase stability and temporal-mode control are active engineering issues rather than trivial assumptions \citep{Bouchard2024,Montaut2025}.

\subsection{Fusion, Bell Measurements, and Temporal-Mode Synchronization}

The recursive fusion process assumed in this framework is idealized.

The current work assumes that Bell-pair sources, delay lines, polarized beam splitters, and Bell-type projections can be composed recursively to generate a growing temporal GHZ chain. This assumption is motivated by the resource-efficient temporal multiphoton entanglement generation demonstrated in \citep{megidish2012resource}. However, the present blockchain construction additionally requires phase-encoded logical data to survive recursive fusion.

This requirement is stronger than the setting in \citep{megidish2012resource}, because the proposed framework requires not only successful temporal GHZ generation but also preservation of the encoded relative phase used for block validation.

The tomography of inductively generated multiphoton states in \citep{megidish2017} supports the relevance of phase-coherent temporal multiphoton descriptions, but it does not by itself enable establishing that phase-encoded blockchain data can be preserved under recursive fusion.

The following assumptions are therefore required:

\begin{itemize}
\item high indistinguishability of photons from different temporal modes;
\item stable temporal overlap at each fusion step;
\item sufficiently high Bell-measurement or fusion success probability;
\item preservation of the encoded relative phase during recursive projection;
\item temporal coherence between non-coexisting time modes;
\item synchronization between quantum-state distribution, validator measurement, and classical consensus messages;
\item an operational phase reference against which relative phase values can be defined and verified.
\end{itemize}

The synchronization assumption is related to recent heralded photonic GHZ-state generation work such as \citep{cao2024}, where temporal control and heralding are central experimental requirements. However, the proposed blockchain requires these ingredients to be extended to a recursively growing ledger state.

These assumptions are nontrivial. Linear-optical Bell-state measurements are generally probabilistic and resource-constrained, and recent photonic-fusion work continues to treat Bell-measurement success probability as a central scalability bottleneck \citep{Arenskoetter2024,Hauser2025}.

Therefore, the fusion process in the present manuscript should be treated as an idealized architectural primitive, not as an experimentally solved blockchain operation.

\subsection{Decoherence, Noise, Entanglement Fidelity, and Network Realization}

The proposed architecture assumes that temporal entanglement can be maintained long enough for block verification and recursive chain extension.

This requires:

\begin{itemize}
\item low decoherence during temporal delay;
\item low photon loss;
\item low phase noise;
\item stable quantum memory and delay-line behavior;
\item high-fidelity entangled-pair generation;
\item high-fidelity multiphoton GHZ-state preparation;
\item reliable authenticated classical side information;
\item quantum-network synchronization across validators.
\end{itemize}

The assumption of negligible decoherence is strong. Coherent control of large quantum states remains experimentally difficult even in advanced platforms; for example, \citep{Omran2019} demonstrates controlled generation and manipulation of large Schrödinger-cat states, but also illustrates that such coherence is a demanding physical resource rather than a default condition.

Similarly, \citep{Xing2023} demonstrates preparation of multiphoton high-dimensional GHZ states, but the present framework assumes a stronger operational requirement: such high-fidelity GHZ-type resources must be generated, phase encoded, distributed, verified, and recursively fused as part of a blockchain data structure.

Existing quantum-network and photonic-platform demonstrations support individual ingredients, such as time-bin encoding, temporal entanglement, photonic GHZ-state generation, and entanglement links. However, they do not establish the feasibility of a recursively growing phase-encoded temporal GHZ ledger \citep{Knaut2024,Bouchard2024,Montaut2025}.

Thus, the proposed blockchain should be interpreted as a conceptual architecture. Its main contribution is the combination of phase-encoded classical data with temporal entanglement, while practical deployment requires future work on error tolerance, phase drift, fusion success probability, decoherence thresholds, entanglement fidelity, external phase referencing, and synchronization.

\section{Limitations}

It must be understood that this work is to merely convey a conceptual blockchain (ledger data structure) on a quantum register. Low-level, case-by-case attacker and security models, a quantum consensus algorithm tailored to the phased GHZ and Bell states in question must be constructed. This work is an initial formulation and not a complete theoretical framework or an engineering-ready design of a blockchain.

\begin{enumerate}
    \item \textbf{Incomplete consensus protocol formalization.} While we outline a high-level consensus mechanism, a fully detailed quantum consensus protocol with rigorous message flows, fault tolerance proofs, and an explicit adversarial model in a quantum network remains to be done. We encourage further research to formalize and extend this framework's basis for consensus. Similar to \citep{rajan2019}, the starting point for this development could consider the \(\theta\)-protocol \citep{McCutcheon2016}.
    
    \item \textbf{Adversary and threat model.} The security analyses assume an idealized quantum network, limited error models, and simplified adversarial capabilities. A comprehensive, case-by-case attacker model (including quantum memory attacks, denial-of-service, network partitioning, and measurement coercion) is not fully treated. 
    
    \item \textbf{Technology and implementation readiness.} The proposal leverages time entanglement, phase encoding, and phased GHZ states. No empirical prototype or detailed engineering study on current photonic hardware, error rates, or scalability in practical deployment is included. Several constituent ingredients underlying the proposal have experimental or simulation-based precedents in prior literature, including temporal entanglement demonstrations and phase-encoding constructions \citep{rajan2019,shreya2020}. However, the integrated phased-temporal GHZ blockchain architecture proposed here has not been experimentally validated. In particular, feasibility at scale, noise tolerance, synchronization constraints, and precise phase control in recursive fusion processes remain open engineering questions.
    
    \item \textbf{Scalability and resource analysis.} While combining time-entanglement with phase encoding aims to enhance efficiency of the temporal GHZ blockchain \citep{rajan2019}, detailed quantitative resource estimates (e.g., qubit counts, entanglement generation overhead, decoherence thresholds, throughput/latency metrics) for realistic network sizes are not provided by us.

    \item \textbf{Fusion process using PBS} In \citep{Megidish_2013}, they successfully produced GHZ states by performing Bell projection on photons from different Bell pairs using polarized beam splitting. That demonstration did not explicitly incorporate the phase-control requirements assumed in the present framework. Extending such fusion processes to maintain precise and scalable phase control remains an open experimental challenge.
    
    \item \textbf{Classical-quantum interface and operational integration.} The transition between classical data and the quantum blockchain register is described conceptually, but a full operational specification (including classical-quantum interaction layers, transaction validation mechanisms, revocation models, and hybrid network protocols) is left for future work.
\end{enumerate}

\section{Future Directions}

Several directions remain open for extending and refining the present framework. First, a rigorous mathematical treatment of phased temporal GHZ states is needed, including explicit analysis of coherence preservation, phase accumulation, and disturbance under realistic noise and measurement models. While temporal entanglement has been experimentally demonstrated in related settings \citep{Megidish_2013,megidish2012resource,megidish2017}, its use as a blockchain data structure with controlled phase encoding requires further theoretical and experimental validation. Also, the proposed consensus mechanism should be formalized into a complete protocol with explicit message complexity, fault tolerance guarantees, adversarial assumptions, and security proofs. Connections to multipartite quantum network verification protocols such as the \(\theta\)-protocol may provide a useful starting point \citep{McCutcheon2016}. In addition, practical feasibility studies are required for implementation over emerging quantum networks. This includes resource estimation for entanglement generation, synchronization across temporal modes, quantum memory requirements, communication overhead, and resilience to decoherence in realistic photonic or hybrid quantum architectures \citep{Simon2017,manssor2024,jiang2024}. Finally, broader conceptual applications of temporally entangled ledgers may be explored. Since temporal entanglement connects non-coexisting quantum events, future work may investigate whether such structures offer new perspectives for secure time-ordered records, causality-aware distributed systems, or information-theoretic studies of temporal quantum correlations \citep{lloyd2011_1,lloyd2011_2,olsen2011,olsen2012,sabin2012}.

\section{Conclusion}

This work proposed a conceptual hybrid quantum blockchain architecture combining phase-based representation of classical block data\citep{shreya2020} with temporal GHZ entanglement\citep{rajan2019, rajan2020quantum}. The main contribution is the integration of phase encoding, temporal entanglement, and verification definitions into a single framework for exploring tamper-evident quantum ledger possibilities. The proposed framework suggests that temporal entanglement and phase-encoded Bell states and GHZ states may provide a route toward disturbance-detectable storage and verification of classical block records. However, the present work does not claim a complete blockchain security proof, unconditional immutability, or experimentally demonstrated scalability. Several important issues remain open. These include formal safety and liveness proofs, malicious-validator analysis, network-level attacks, decentralized quantum public-key distribution, hardware-specific noise models, and detailed comparison with classical cryptographic blockchain mechanisms. Therefore, the framework should be interpreted as a conceptual step toward quantum-assisted tamper-evident ledgers rather than a deployable secure blockchain protocol.


\bibliographystyle{unsrt}  






\end{document}